\documentclass{sf2a-conf2011}
\usepackage{graphicx}
\usepackage{hyperref}
\usepackage[]{natbib}
\usepackage[cyr]{aeguill}
\usepackage{epstopdf}
\usepackage{amsmath}
\usepackage{amsfonts}
\usepackage{cases}

\def\BibTeX{{\rm B\kern-.05em{\sc i\kern-.025em b}\kern-.08em
    T\kern-.1667em\lower.7ex\hbox{E}\kern-.125emX}}
\bibpunct{(}{)}{;}{a}{}{,}  

\newcommand{\ve}{\mathbf}

\begin{document}

\TitreGlobal{SF2A 2011}


\title{Measuring the absolute non-gravitational acceleration of a spacecraft: goals, devices, methods, performances}
\runningtitle{Measuring the  absolute non-gravitational acceleration of a spacecraft}
\author{B. Lenoir}\address{Onera -- The French Aerospace Lab, 29 avenue de la Division Leclerc, F-92322 Ch\^atillon, France.}
\author{B. Christophe$^1$}
\author{S. Reynaud}\address{Laboratoire Kastler Brossel (LKB), ENS, UPMC, CNRS, Campus Jussieu, F-75252 Paris Cedex 05, France.}

\setcounter{page}{1}

\index{Lenoir, B.}
\index{Christophe, B.}
\index{Reynaud, S.}


\maketitle


\begin{abstract}
Space provides unique opportunities to test gravitation. By using an interplanetary spacecraft as a test mass, it is possible to test General Relativity at the Solar System distance scale. This requires to compute accurately the trajectory of the spacecraft, a process which relies on radio tracking and is limited by the uncertainty on the spacecraft non-gravitational acceleration.

The Gravity Advanced Package (GAP) is designed to measure the non-gravitational acceleration without bias. It is composed of an electrostatic accelerometer supplemented by a rotating stage. This article presents the instrument and its performances, and describes the method to make unbiased measurements. Finally, it addresses briefly the improvement brought by the instrument as far as orbit reconstruction is concerned.
\end{abstract}

\begin{keywords}
Electrostatic accelerometer, Bias rejection, Colored noise, Allan variance, Non-gravitational acceleration, General Relativity, Orbit reconstruction
\end{keywords}


\section{Introduction}

With the ever-increasing precision of measurements, space has become a privileged place to test the two fundamental theories which emerged during the 20\textsuperscript{th} century: General Relativity and Quantum Theory. In addition to providing a very clean environment, it opened new ways of testing these theories: as an example of interest for this article, precise navigation of interplanetary spacecrafts allows probing the scale dependence of gravitation at the Solar System distance scale \citep{jaekel2006radar}.

Even if most experimental tests support General Relativity \citep{will2006confrontation}, there are still open windows for deviations. Indeed, the fact that these two fundamental theories are difficult to reconcile suggest that General Relativity may not be the final description of gravitation. The reason is that gravitation is the only interaction not having a quantum description. The validity of the Newton potential has been extensively tested for distances between the millimeter and the characteristic size of planetary orbits \citep{fischbach1999search}. But there remain open windows outside this distance range for violations of the inverse square law \citep[][Fig.~4]{adelberger2003tests}: below the millimeter or for distances of the order or larger than the Solar System characteristic size.

Long range tests are performed using the motion of planets and interplanetary probes. Monitoring of the Moon and Mars delivers high precision tests of the validity of General Relativity at these distances \citep[e.g.][]{kolosnitsyn2004test,williams1996relativity}. However, the navigation data of the Pioneer probes show a discrepancy with respect to the predictions of General Relativity \citep{anderson1998indication,anderson2002study,levy2009pioneer}. This discrepancy can be described as an anomalous acceleration directed toward the Sun with a roughly constant amplitude of approximately $8 \times 10^{-10}$~m.s$^{-2}$. The origin of this anomaly is yet unexplained despite a huge effort of the scientific community \citep[][and references therein]{turyshev2010pioneer}: it may be an experimental artifact as well as a hint of considerable importance for fundamental physics \citep{brownstein2006gravitational,jaekel2005gravity}. At larger scales, the rotation curves of galaxies and the relation between redshift and luminosities of supernovae are accounted for by introducing respectively ``dark matter'' and ``dark energy'', which represent 25~\% and 70~\% of the energy content of the Universe \citep{frieman2008dark}. Since these dark components have been introduced on the basis of gravitational observations solely, the hypothesis that General Relativity is not a correct description of gravitation at these scales needs to be considered \citep{carroll2004cosmic}.

It is therefore essential to test gravitation at all distance scales. To this extend, several mission concepts have been proposed to improve the experiment made by the Pioneer probes \citep{anderson2002mission,bertolami2007mission,johann2008exploring,christophe2009odyssey,christophe2011oss,wolf2009quantum}. In many proposals, the addition of an accelerometer being able to measure without bias the non-gravitational acceleration of the spacecraft is central. ESA included this idea in the roadmap for fundamental physics in space \citep{esa2010roadmap} and recommended the development of accelerometer compatible with spacecraft tracking at the 10 pm.s$^{-2}$ level. This article presents such an instrument, called the Gravity Advanced Package \citep{lenoir2011electrostatic}. First, a description of the instrument and its performances is given. Then, the method used to make absolute measurement is described \citep{lenoir2011unbiased}. Finally, the expected improvements of the orbit reconstruction process using the instrument are briefly discussed.

\section{The Gravity Advanced Package}

The Gravity Advanced Package is an important technological upgrade for future fundamental physics missions in space. It is composed of two subsystems: MicroSTAR is a three-axis electrostatic accelerometer \citep{josselin1999capacitive} based on Onera's experience \citep{touboul1999electrostatic,touboul2001microscope}, and the Bias Rejection System is a rotating stage with piezo-electric actuator used to rotate MicroSTAR around its $x$~axis. The accelerometer aims at measuring the non-gravitational acceleration of the spacecraft but other quantities are also measured. In fact, MicroSTAR measures the components of the vector $\ve{a}$ on its three orthogonal measurement axis called $x$, $y$ and $z$ :
\begin{equation}
  \ve{a}  =  \frac{1}{m_S}\ve{F^{NG}_{\mathnormal{ext} \rightarrow \mathnormal{S}}} + \ve{\dot{\Omega}} \wedge\ve{l} + \ve{\Omega} \wedge \left(\ve{\Omega} \wedge\ve{l} \right) - \left( \frac{1}{m_A} + \frac{1}{m_S} \right) \ve{F^{G}_{\mathnormal{S}\rightarrow \mathnormal{A}}} + \left(\frac{1}{m_S} \ve{F^{G}_{\mathnormal{ext}\rightarrow \mathnormal{S}}}  -  \frac{1}{m_A} \ve{F^{G}_{\mathnormal{ext}\rightarrow \mathnormal{A}}}\right)
  \label{eq:dynamics_final}
\end{equation}
where $m_S$ and $m_A$ are the masses of the satellite and the proof mass respectively, $\ve{F^{NG}_{\mathnormal{ext} \rightarrow \mathnormal{S}}}$ is the non-gravitational force acting on the spacecraft, $\ve{\Omega}$ is the rotation vector of the instrument with respect to a Galilean reference frame, $\ve{l}$ is the vector between the center of mass of the satellite and the instrument (lever arm), $\ve{F^{G}_{\mathnormal{S}\rightarrow \mathnormal{A}}}$ is the gravity of the spacecraft, and the last term in parenthesis is the gravity gradient expressed in term of acceleration, $\ve{F^{G}_{\mathnormal{ext}\rightarrow \mathnormal{S}}}$ and $\ve{F^{G}_{\mathnormal{ext}\rightarrow \mathnormal{A}}}$ being the gravitational forces acting on the satellite and the proof mass respectively. All these additional terms can be removed \citep{lenoir2011electrostatic}.

Of course, the measurement is plagued by scale ($\delta k_{1\nu}$) and quadratic factors ($k_{2\nu}$), by bias ($b_\nu$) and by noise ($n_\nu$), such that the actual measurement on the axis $\nu\in\{x;y;z\}$ is :
\begin{equation}
 m_\nu = (1+\delta k_{1\nu}) a_\nu + k_{2\nu} {a_\nu}^2+b_\nu+n_\nu
\end{equation}
where $a_\nu$ is the projection of $\ve{a}$ on the axis $\nu$.

\begin{figure}[ht!]
 \centering
 \includegraphics[width=0.48\textwidth,clip]{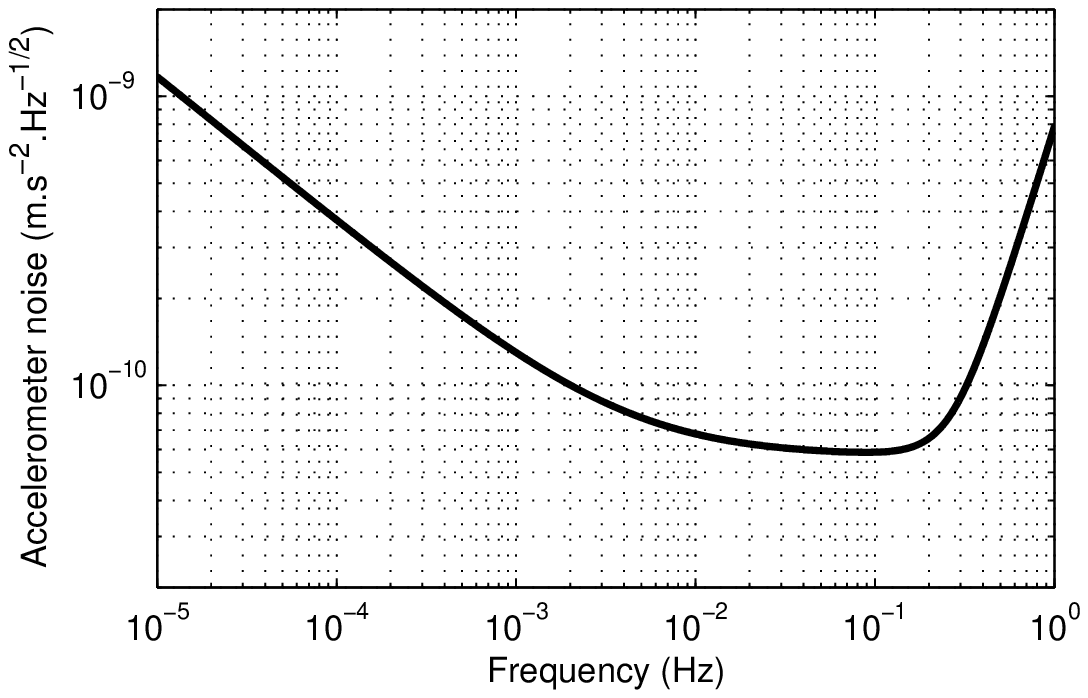}
 \includegraphics[width=0.48\textwidth,clip]{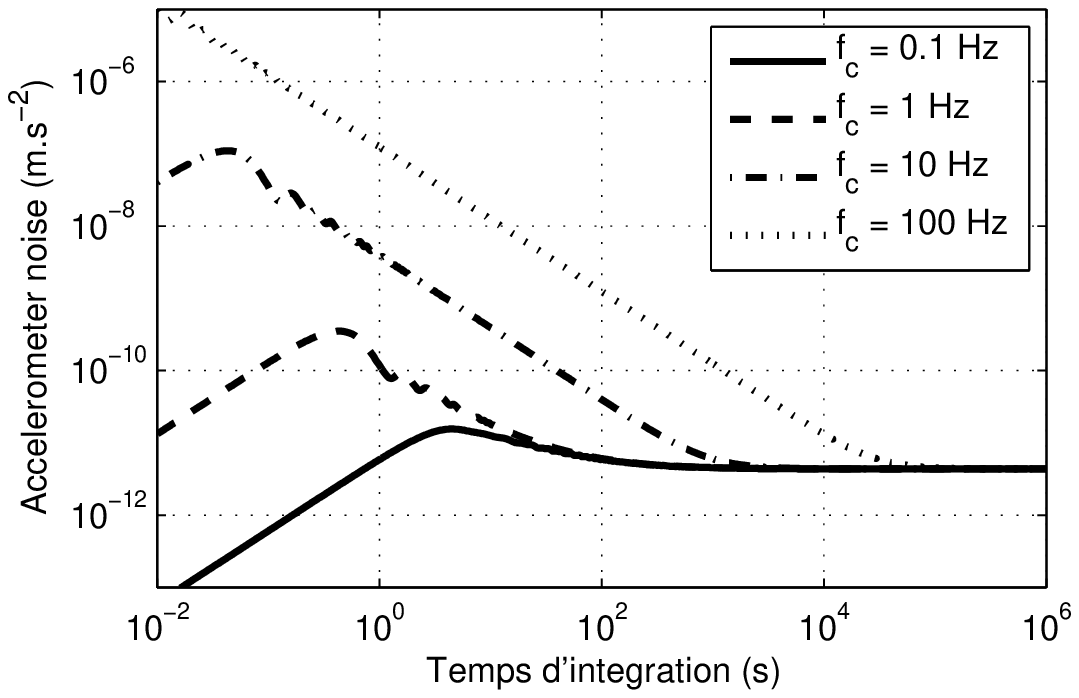}
  \caption{{\bf Left:} Square-root of the power spectrum density  of the accelerometer noise. {\bf Right:} Square-root of the Allan variance of the accelerometer noise. The curves are plotted for different cut-off frequencies. The oscillations for $2\pi\tau f_c < 1$ are not described by the simplified formula \eqref{eq:allan}.}
  \label{fig:noise}
\end{figure}

The measurement noise is characterized by the following power spectrum density \citep{lenoir2011electrostatic} for a measurement range of $1.8 \times 10^{-4}$ m.s$^{-2}$ (cf. fig. \ref{fig:noise}):
\begin{equation}
  S_{n}(f) = \left(5.7 \times 10^{-11} \ \mathrm{m}.\mathrm{s}^{-2}.\mathrm{Hz}^{-1/2}\right)^2 \times \left[ 1 + \left(\frac{f}{4.2 \ \mathrm{mHz}}\right)^{-1} + \left(\frac{f}{0.27 \ \mathrm{Hz}}\right)^{4} \right]
  \label{eq:acc_noise}
\end{equation}
The characterization of the noise can also be given in term of Allan variance $A(\tau,f_c)$, where $f_c$ is the cut-off frequency. A simplified expression of the Allan variance (AVAR) \citep{allan1966statistics} for $2\pi\tau f_c \gg 1$ is:
\begin{equation}
 A_n(\tau,f_c) =  \left(5.7 \times 10^{-11} \ \mathrm{m}.\mathrm{s}^{-2}.\mathrm{Hz}^{-1/2}\right)^2 \times \left[ \frac{1}{2\tau} + 2 \ \mathrm{ln}(2) \times 4.2 \ \mathrm{mHz}  + \frac{f_c^3}{3\pi^2\tau^2} \times (0.27 \ \mathrm{Hz})^{-4} \right]
 \label{eq:allan}
\end{equation}
Figure \ref{fig:noise} shows the dependence of Allan variance with respect to integration time and cut-off frequency without the $2\pi\tau f_c \gg 1$ approximation.

The Bias Rejection System rotates the accelerometer around the $x$ axis of a monitored angle called $\theta$. Assuming that the quadratic factors are equal to zero and that the scale factors are perfectly known\footnote{These assumptions are made for simplicity purpose. The complete treatment of the problem is presented in \citep{lenoir2011unbiased}.}, the quantities measured along the axis $y$ and $z$ are :
\begin{subequations}
  \begin{numcases}{}
    m_y = \left[\cos(\theta) a_Y  + \sin(\theta) a_Z \right] + b_y + n_y \\
    m_z = \left[-\sin(\theta) a_Y + \cos(\theta) a_Z \right] + b_z + n_z
  \end{numcases}
  \label{eq:measurement}
\end{subequations}
with $a_\mu$ ($\mu\in \{Y;Z\}$) being the components of the acceleration in the reference frame of the spacecraft.

\section{Method to remove the bias of the electrostatic accelerometer from the measurements}

The method for removing the bias of the instrument consists in flipping MicroSTAR. The underlying idea is that when $\theta=0$~rad, the accelerometer measures the quantities $m_y = a_Y+b_y$ and $m_z = a_Z+b_z$, and when $\theta=\pi$~rad, it measures  $m_y = -a_Y+b_y$ and $m_z = -a_Z+b_z$. Subtracting these measurements allows recovering the external signal without bias, under the assumption that they are constants. The complete method, which can handle time variations of the external signal and of the bias, is described in \citep{lenoir2011unbiased}. It is shown in particular that the modulation signal, i.e. the time variation of $\theta$, must fulfill some conditions in order to correctly remove the bias from the measurement (cf. eq. \eqref{eq:condition}).

The modulation signal is supposed to be periodic, $\tau$ being the period. Moreover, the measurements used for data post-processing are the ones made when the angle $\theta$ is constant with time and only two positions are considered: $0$~rad and $\pi$~rad (so that assuming $k_{2\nu}=0$ is not restrictive). On the contrary, the measurements made when the accelerometer is rotating are not used because they may be spoiled by unwanted signal (vibration, fictitious acceleration). The duration of the rotation per period is called $T_M$ and will be referred to as the masking duration.

\begin{figure}[ht!]
 \centering
 \includegraphics[width=0.48\textwidth,clip]{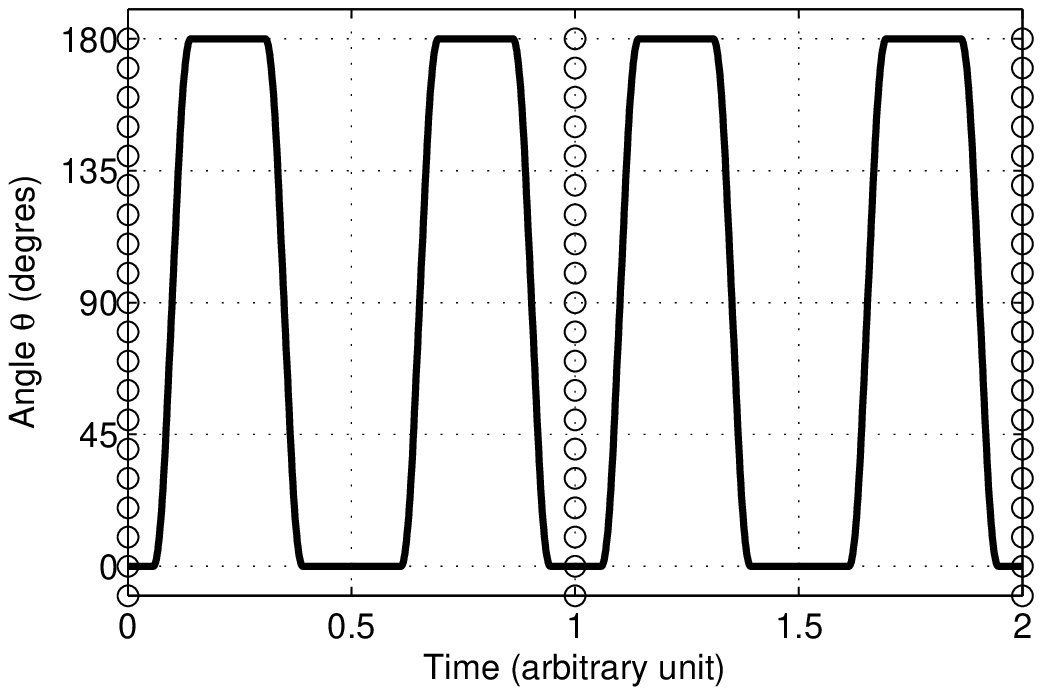}
 \includegraphics[width=0.48\textwidth,clip]{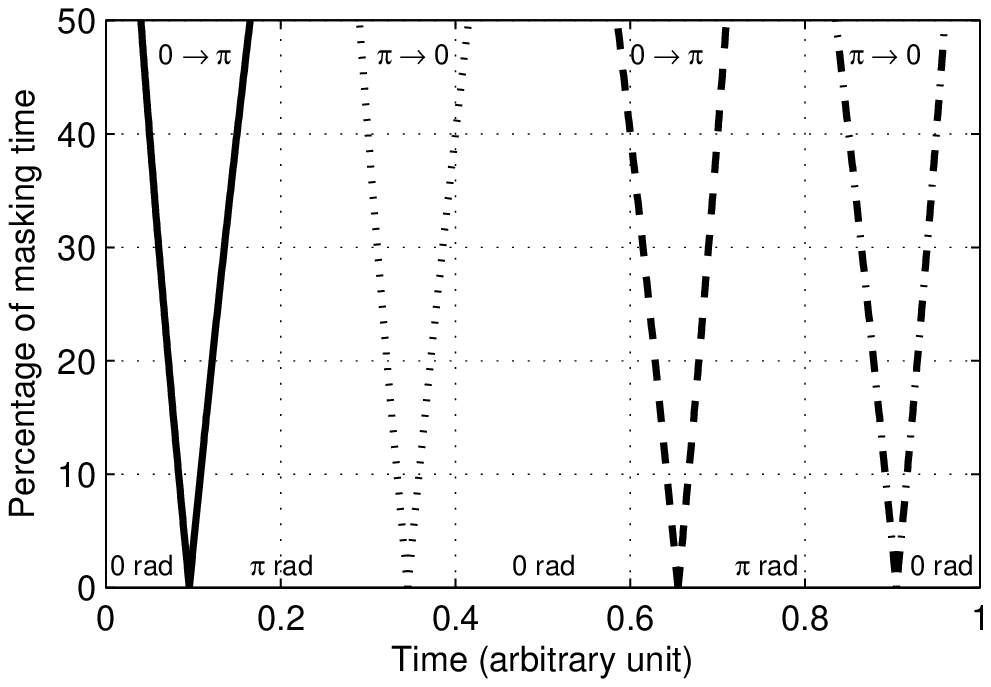}
  \caption{{\bf Left:} Modulation signal (-) for a masking duration representing 33.33~\% of the modulation period $\tau = 1$~arbitrary unit. Two periods are represented, separated by circles ($\circ$) {\bf Right:} The figure describes how the modulation signal changes when the ratio $T_M/\tau$ changes (one period is represented). The abscisse is time and the ordinate is $T_M/\tau$ expressed in percentage. The curves indicate the values of the angle $\theta$: between two curves of the same style $\theta$ changes, between the curves (-) and ($\cdot$) and between the curves (- -) and (- $\cdot$) $\theta=\pi$~rad, and $\theta=0$~rad elsewhere.}
 \label{fig:lin_signal}
\end{figure}

Assuming that the signal to measure and the bias are affine functions of time for each modulation period (which is correct if $\tau$ is small compared to the characteristic variation time of the signal and the bias), the conditions for completely removing the bias from the measurements read
\begin{equation}
 \int_{-\tau/2}^{\tau/2} m(t) \cos(\theta(t)) dt =  \int_{-\tau/2}^{\tau/2} t \ m(t) \cos(\theta(t)) dt =  \int_{-\tau/2}^{\tau/2} t^2 \ m(t) \cos(\theta(t)) dt = 0
 \label{eq:condition}
\end{equation}
where $m(t)$ is equal to $0$ when the accelerometer is rotating and $1$ when it is not. These conditions allow deriving the time variation of the angle $\theta$, which is shown in figure \ref{fig:lin_signal} (left). The pattern depends on the ratio of the masking duration $T_M$ and the period $\tau$, as shown by figure \ref{fig:lin_signal} (right).

With such a signal and after post-processing, it is possible to recover the mean of the external signal without bias over a modulation period $\tau$. It is possible to characterize these unbiased measurements in term of noise. The level of noise depends on the modulation period $\tau$ and the masking time $T_M$ but is approximately white as shown by figure \ref{fig:dsp} (left). For $\tau=600$~s and $T_M=200$~s, the uncertainty on the unbiased measurements is $4.2 \times 10^{-12}$~m.s$^{-2}$ (this value is the integral of the power spectrum density shown in fig.~\ref{fig:dsp} (left)). This allows reaching a precision of $1$~pm.s$^{-2}$ for an integration time of three hours.

\section{Improvement of orbit reconstruction}

As mentioned in the introduction, the Gravity Advances Package (GAP) aims at improving the orbit reconstruction of interplanetary probes so as to test the theories of gravitation. So far, models have been used to correct for the non-gravitational acceleration of the spacecrafts. But the computed orbit is then subject to errors due to uncertainties or inaccuracies in the models. By providing unbiased measurements of the non-gravitational acceleration in the orbit plane, the GAP enhances orbit reconstruction: it removes parameters to be fitted, it measures the fluctuation of the non-gravitational acceleration, and it removes correlations.

\begin{figure}[ht!]
 \centering
 \includegraphics[width=0.48\textwidth,clip]{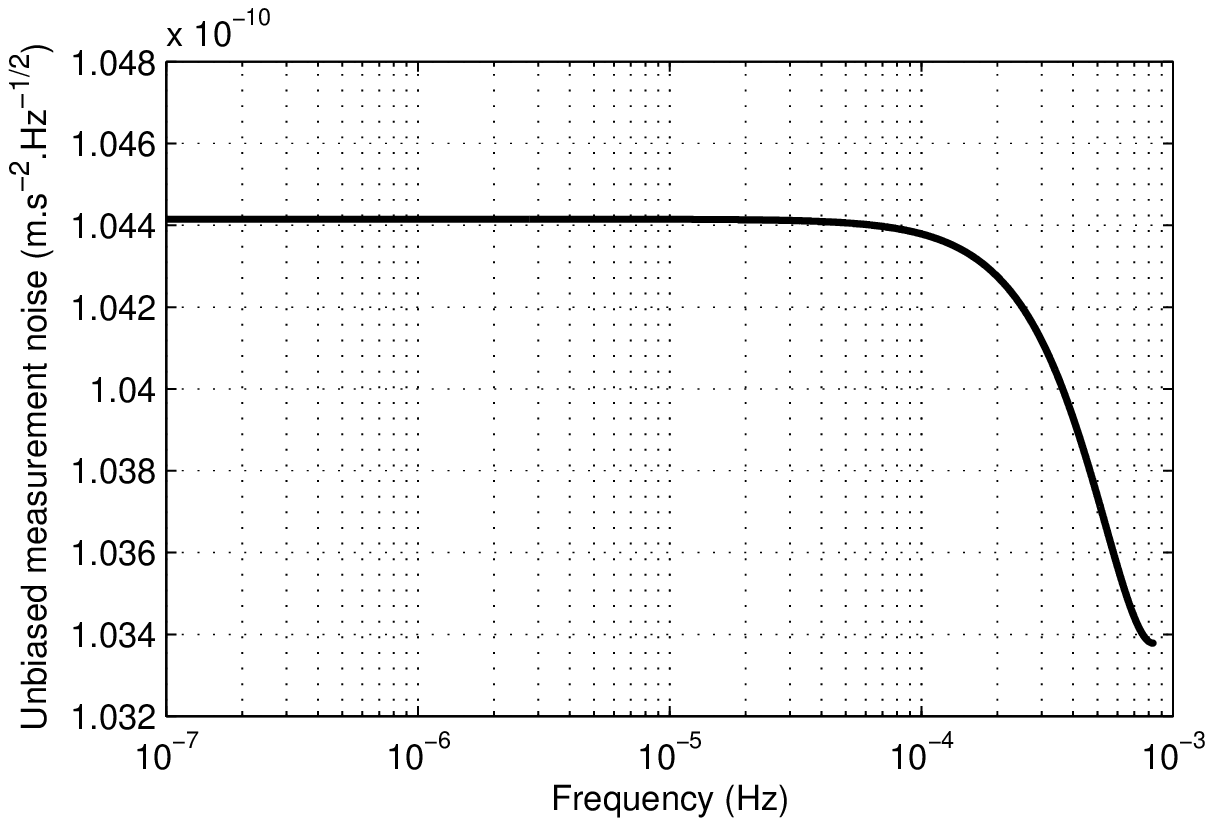}
 \includegraphics[width=0.48\textwidth,clip]{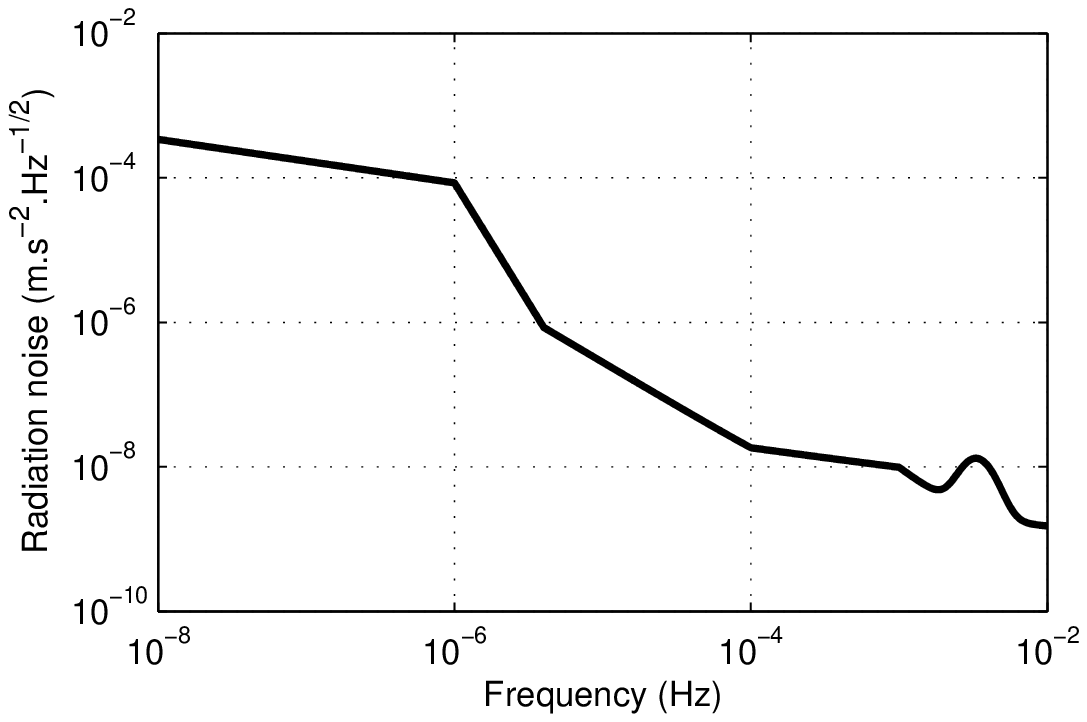}
  \caption{{\bf Left:} Square root of the power spectrum density of the unbiased measurements obtained after post-processing for a modulation period $\tau=600$~s and a masking duration $T_M=200$~s. {\bf Right:} Square root of the power spectrum density of the solar radiation pressure noise \citep{frohlich2004solar} in term of acceleration for a spacecraft with a ballistic coefficient of 0.1~m$^2$.kg$^{-1}$ at 10 AU of the Sun \citep{biesbroek2008laplace}. The bump at $3$~mHz corresponds to the 5-minute oscillations of the Sun.}
  \label{fig:dsp}
\end{figure}

There are several sources for the non-gravitational acceleration, the main ones being the direct solar radiation pressure and the anisotropic thermal radiation of the spacecraft. Figure \ref{fig:dsp} illustrates the interest of the GAP as far as the fluctuations of the non-gravitational acceleration are concerned. The power spectrum density of the unbiased measurement noise is compared to the noise on the radiation pressure expressed in term of acceleration. Whereas these fluctuations are not taken into account in the models, it shows that the GAP allows measuring them with a very high precision.

\section{Conclusions}

The Gravity Advances Package, designed to improve the tests of gravitation at the Solar System distance scale, displays performances which will allow improving the accuracy of orbit reconstruction significantly. Indeed, with a carefully designed calibration signal for the rotating stage, it is possible to remove completely the bias of the electrostatic accelerometer and to obtain the mean non-gravitational acceleration of the spacecraft with no bias and a precision of 1 pm.s$^{-2}$ for an integration time of three hours. This expected precision will allow reaching the 10 pm.s$^{-2}$ level for spacecraft tracking recommended by ESA. To do so, the accelerometer measurements will have to be taken into account during the orbit reconstruction process.

\begin{acknowledgements}
The authors are grateful to CNES for its financial support.
\end{acknowledgements}


\end{document}